\documentclass[preprint,12pt,a4paper]{elsarticle}

\usepackage{amssymb}
\usepackage{ntheorem}
\theoremstyle{break}
\newtheorem{definition}{Definition}[section]
\theoremstyle{break}
\newtheorem{theorem}{Theorem}[section]

\usepackage{algpseudocode}
\usepackage{algorithm}

\usepackage{lineno}
\usepackage{graphicx,amsmath,amsfonts,amssymb,bm,bigints}
\usepackage{hyperref}
\usepackage[nameinlink,capitalize]{cleveref}
\usepackage{subfigure}
\usepackage{hhline}
\usepackage{booktabs}
\usepackage{multirow}
\usepackage{listings}
\usepackage{enumitem}
\usepackage{comment}

\usepackage{epstopdf}
\usepackage[colorinlistoftodos]{todonotes} 
\usepackage{xcolor,colortbl}
\usepackage[english]{babel}

\newcommand{\tablistcommand}{%
    \leavevmode\par\vspace{-\baselineskip}%
}
\newlist{tabitemize}{itemize}{1}
\setlist[tabitemize]{%
    leftmargin = *               ,
  label      = \textbullet     ,
  nosep                        ,
  before     = \tablistcommand ,
  after      = \tablistcommand
}

\newcommand{\vecx}{{\bf x}}
\newcommand{\vecy}{{\bf y}}

\newcommand{\vecsigma}{{\bm \sigma}}
\renewcommand{\O}{\mathcal{O}}

\definecolor{Gray}{gray}{0.85}

\begin{document}

\begin{frontmatter}



\title{\texttt{PBBFMM3D}: a parallel black-box algorithm for kernel matrix-vector multiplication}



\author[add1]{Ruoxi Wang}
\ead{ruoxi.rw@gmail.com}
\author[add1]{Chao Chen}
\ead{chenchao.nk@gmail.com}
\author[add2]{Jonghyun Lee}
\ead{jonghyun.harry.lee@hawaii.edu}
\author[add1,add3]{Eric Darve}
\ead{darve@stanford.edu}

\address[add1]{Institute for Computational and Mathematical Engineering, Stanford University}
\address[add2]{Department of Civil and Environmental Engineering \& Water Resources Research Center, University of Hawai'i at M\=anoa}
\address[add3]{Department of Mechanical Engineering, Stanford University}

\begin{abstract}
Kernel matrix-vector product is ubiquitous in many science and engineering applications. However, a naive method requires $\O(N^2)$ operations, which becomes prohibitive for large-scale problems. To reduce the computation cost, we introduce a parallel method that provably requires $\O(N)$ operations and delivers an approximate result within a prescribed tolerance. The distinct feature of our method is that it requires only the ability to evaluate the kernel function, offering a black-box interface to users. Our parallel approach targets multi-core shared-memory machines and is implemented using  \verb|OpenMP|. Numerical results demonstrate up to $19\times$ speedup on 32 cores. We also present a real-world application in geo-statistics, where our parallel method was used to deliver fast principle component analysis of covariance matrices.
\end{abstract}


\end{frontmatter}


\section{Introduction}
\label{sec:intro}

We consider the problem of computing kernel matrix-vector products, 
where the kernel function $\mathcal{K}(\vecx, \vecy)$ is \textbf{n}on-\textbf{o}scillatory, \textbf{t}ranslation \textbf{i}nvariant, and sufficiently \textbf{s}mooth (NOTIS) ($\mathcal{K}$ can be singular when $\vecx=\vecy$). For example, a NOTIS function can be $1/\|\vecx - \vecy\|$ or exp($-\|\vecx - \vecy\|$).

The problem can be formulated mathematically as evaluating
\begin{equation} \label{eq:summation}
\phi_i = \sum_{j=1}^N \mathcal{K}(\vecx_i, \vecy_j) \,\, \sigma_j, \quad i = 1, \ldots, N
\end{equation}
where $\{\vecx_i\}_{i=1}^N$ and $\{\vecy_i\}_{i=1}^N$ are the target and the source data points in a cubical domain, respectively, and $\sigma_j$ is the weight associated with $\vecy_j$.
In many applications, the two sets of points $\{\vecx_i\}_{i=1}^N$ and $\{\vecy_i\}_{i=1}^N$ may overlap. Algebraically, \cref{eq:summation} can be written as the following matrix-vector multiplication:
\begin{equation} \label{eq:matvec}
    \bm{\phi} = \bm{K}\vecsigma,
\end{equation}
where $\vecsigma = [\sigma_1, \ldots, \sigma_N]^T$ and $\bm{\phi} = [\phi_1, \ldots, \phi_N]^T$ are two vectors, and $\bm{K}$ is an $N$-by-$N$ matrix with $\bm{K}_{ij} = \mathcal{K}(\vecx_i, \vecy_j)$. Such type of computation arises in many science and engineering fields, such as kernel methods in statistical learning and machine learning~\cite{gray2001n, hofmann2008kernel}, data assimilation methods in geosciences~\cite{ambikasaran2013fast, li2014kalman}, particle simulations and boundary integral/element methods in computational physics~\cite{greengard1987fast, greengard1997new, simpson2016acceleration}, dislocation dynamics simulations in material science~\cite{zhao2012fast, chen2018fast}, etc. To compute $\bm{\phi}$ in \cref{eq:matvec}, a naive direct evaluation requires $\O(N^2)$ operations, which is prohibitive when $N$ is large.

\subsection{Related work}

One special but important instance of \cref{eq:summation} is when the data points $\{\vecx_i\}_{i=1}^N$ and $\{\vecy_i\}_{i=1}^N$ lie on a regular grid. In this case, \cref{eq:summation} can be evaluated exactly (up to round-off errors) using the fast Fourier transform (FFT), which requires  $\O(N \log(N))$ operations. Although the FFT can be extended to handle non-uniform data distributions~\cite{alfke2018nfft, ruiz2018nonuniform}, but its efficiency  decreases for highly irregular distributions in three dimensions (3D).


The fast multipole method (FMM) is a general framework that has been successfully applied to non-uniform data distributions. In the FMM, we partition the problem domain hierarchically into subdomains of different scales and exploit the multi-scale decomposition in the evaluation of \cref{eq:summation}. In particular,  we evaluate exactly the calculation associated with adjacent subdomains at the finest scale; and we evaluate approximately the calculation associated with  non-adjacent subdomains at  every scale. Overall, the procedure requires $\O(N)$ operations with a provable accuracy.
{While the FMM has been derived for specific kernels that appear frequently in computational physics~\cite{greengard1987fast, fu1998fast, fu2000fast, greengard2002new, yoshida2001application}, the derivation may be tedious, difficult or even impossible for an arbitrary   NOTIS kernel function.}

Therefore, black-box algorithms have been developed, which require only the evaluation of a given kernel function. Such methods can be classified into two groups. The first consists of methods that approximate the kernel function (away from the origin) with polynomials, such as Legendre polynomials or Chebyshev polynomials~\cite{dutt1996fast, gimbutas2003generalized, fong2009black, chen2018fast}. The other group consists of methods that compute the so-called equivalent densities or the so-called skeletons for every subdomain to efficiently represent the contained source points and their weights~\cite{ying2004kernel, martinsson2007accelerated, malhotra2015pvfmm, yu2017geometry}. Theoretically, {this approach is justified by the potential theory for kernel functions that are fundamental solutions of non-oscillatory elliptic partial differential equations.}

\subsection{Contributions}
In this paper, we present a parallel implementation of the black-box method in~\cite{fong2009black, chen2018fast} for evaluating \cref{eq:summation} using $\O(N)$ memory and computation. The key idea is using Lagrange interpolation to construct approximations of the kernel function $\mathcal{K}(\vecx, \vecy)$ when $\vecx$ and $\vecy$ are distant. Unlike fast algorithms that have been developed for specific kernel functions, our method applies to a wide range of functions. Successful stories include applications of our method in dislocation dynamics simulations~\cite{chen2018fast} and aquifer characterization~\cite{lee2018fast}. In those two applications, no  fast algorithm exists for the two kernel functions---the Green's function in anisotropic elasticity and the isotropic exponential function. In particular, our method requires only a black-box routine to evaluate the kernel function, and thus can be integrated easily with other codes. For example, our package was used in an iso-geometric boundary element method to obtain a solver of $\O(N)$ complexity~\cite{simpson2016acceleration}. Other examples of using (an earlier version of) our code are in the elastic formulation of the displacement discontinuity method for the simulation of micro-seismicity~\cite{verde2013efficient, verde2015fast, farmahini2016simulation}. Extension of our algorithm for solving/factorizing kernel matrices has been explored in~\cite{coulier2017inverse,takahashi2020parallelization}.


To evaluate \cref{eq:summation}, our method follows the general FMM machinery as follows. First, the problem domain containing $\{\vecx_i\}_{i=1}^N$ and $\{\vecy_i\}_{i=1}^N$ is partitioned in a hierarchical fashion (see \cref{fig:tree}(a)), and the partitioning is associated with a tree data structure where every tree node represents a subdomain in the hierarchy. Second, a post-order traversal of the tree is performed, and the ``multipole coefficients'' associated with every tree node is computed as a compact representation of the source points and corresponding weights that the subdomain contains. This step is often called an upward pass. Third, at every level of the hierarchy, the ``local coefficients'' of every tree node is computed using the ``multipole coefficients'' of its \emph{interaction list} (two nodes are in each other's interaction list if their parents are adjacent but they are not; see \cref{fig:tree}(b)). Fourth, a pre-order traversal of the tree is performed to ``accumulate'' the local coefficients of all nodes to those at the leaf level. This step is often called a downward pass. Finally, the contribution from adjacent subdomains are evaluated exactly for all leaf nodes. Technically speaking, multipole and local coefficients are terminologies used in the original FMM~\cite{greengard1987fast}. Here we adhere to the same terms for their counterparts in our algorithm, which we will rigorously define in~\cref{sec:fmm}.

Our parallel algorithm is based on three  observations of the above procedure: (1) the computation of multipole coefficients and local coefficients during the upward pass and the downward pass is embarrassingly parallel for tree nodes at the same level; (2) the computation of local coefficients based on multipole coefficients  is embarrassingly parallel for all nodes; and (3) the exact evaluation of the contribution from adjacent subdomains is embarrassingly parallel for all leaf nodes. While our parallel algorithm does not exploit task-level parallelism~\cite{agullo2014task}, which usually requires a well-designed runtime system to implement,  our results show satisfactory parallel speedups. 

To summarize, we present {p}arallel {b}lack-{b}ox {FMM} in {3D} (\texttt{PBBFMM3D}) to compute the kernel matrix-vector multiplication in \cref{eq:summation} for NOTIS kernels. The algorithm provably requires $\O(N)$ memory and operations. Our parallel implementation is based on \verb|OpenMP|\footnote{\url{https://www.openmp.org/}} and targets multi-core shared-memory machines. {Nowadays, on-node parallelism is becoming increasingly important because of the wide adoption of multi-core architectures. Our method also serves as the basis for a multi-node distributed-memory solution, which can be implemented on top of our code using the so-called local essential tree and space-filling curve~\cite{warren1992astrophysical,warren1993parallel}. Extensions to many-core architectures (e.g., GPUs) was considered in~\cite{takahashi2012optimization,agullo2016task}, which used blocking schemes to reduce memory movement and to improve the arithmetic intensity (flop-to-word ratio).} Our code is publicly available at 
\begin{center}
\url{https://github.com/ruoxi-wang/PBBFMM3D}    
\end{center}



\subsection{Outline}
The rest of this paper is organized as follows. \cref{sec:algorithm} describes the algorithm focusing on the data dependency and the parallelism. \cref{sec:software_description} describes the software architecture and the user interface.
 \cref{sec:examples} presents the accuracy and the running time of \texttt{PBBFMM3D}. 
 \cref{sec:conclusion} draws the conclusion.

\section{Parallel black-box algorithm} \label{sec:algorithm}
In this section, we present the black-box algorithm for evaluating \cref{eq:summation}. In particular, we focus on the data dependency and illustrate our parallel strategy. Although our code deals with 3D problems, we use 1D and 2D examples here for pictorial illustration.

\subsection{Hierarchical domain decomposition}
\begin{figure}[htbp]
\begin{center}
\subfigure[]{\includegraphics[trim=0 15 0 0, clip, width=5.5cm]{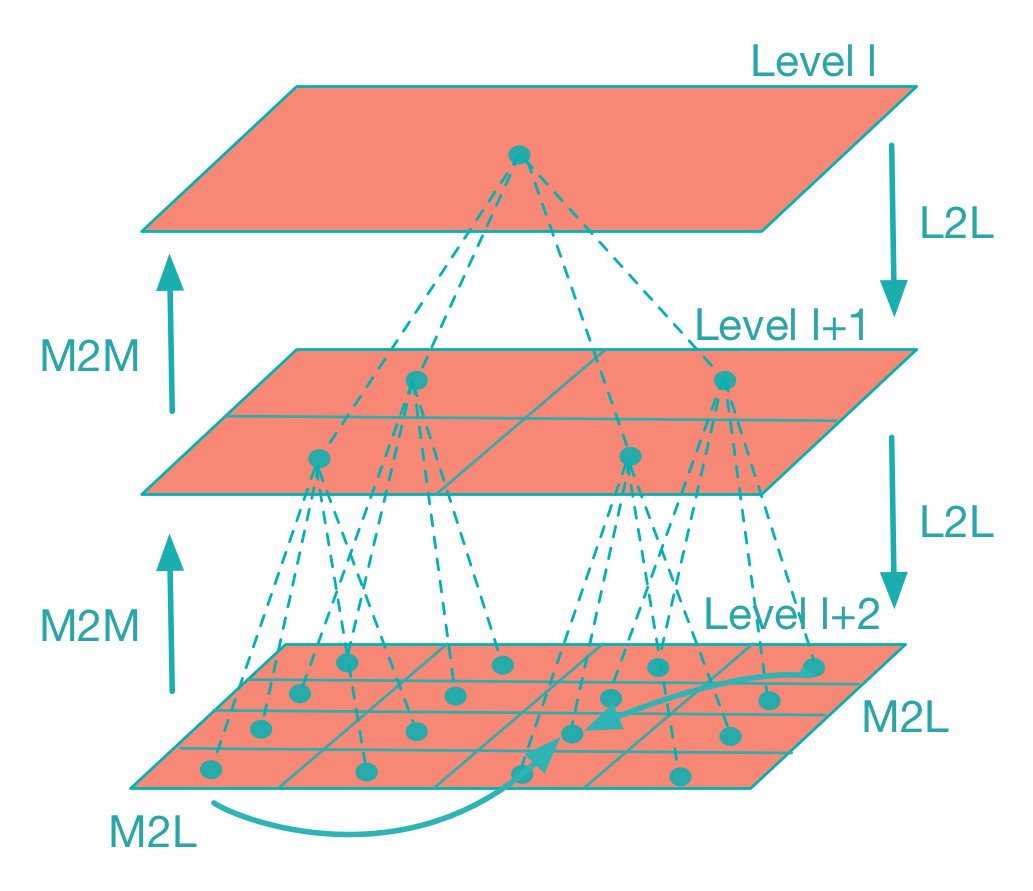}} \hfill
\subfigure[]{\includegraphics[trim=0 0 0 10, clip, width=7.5cm]{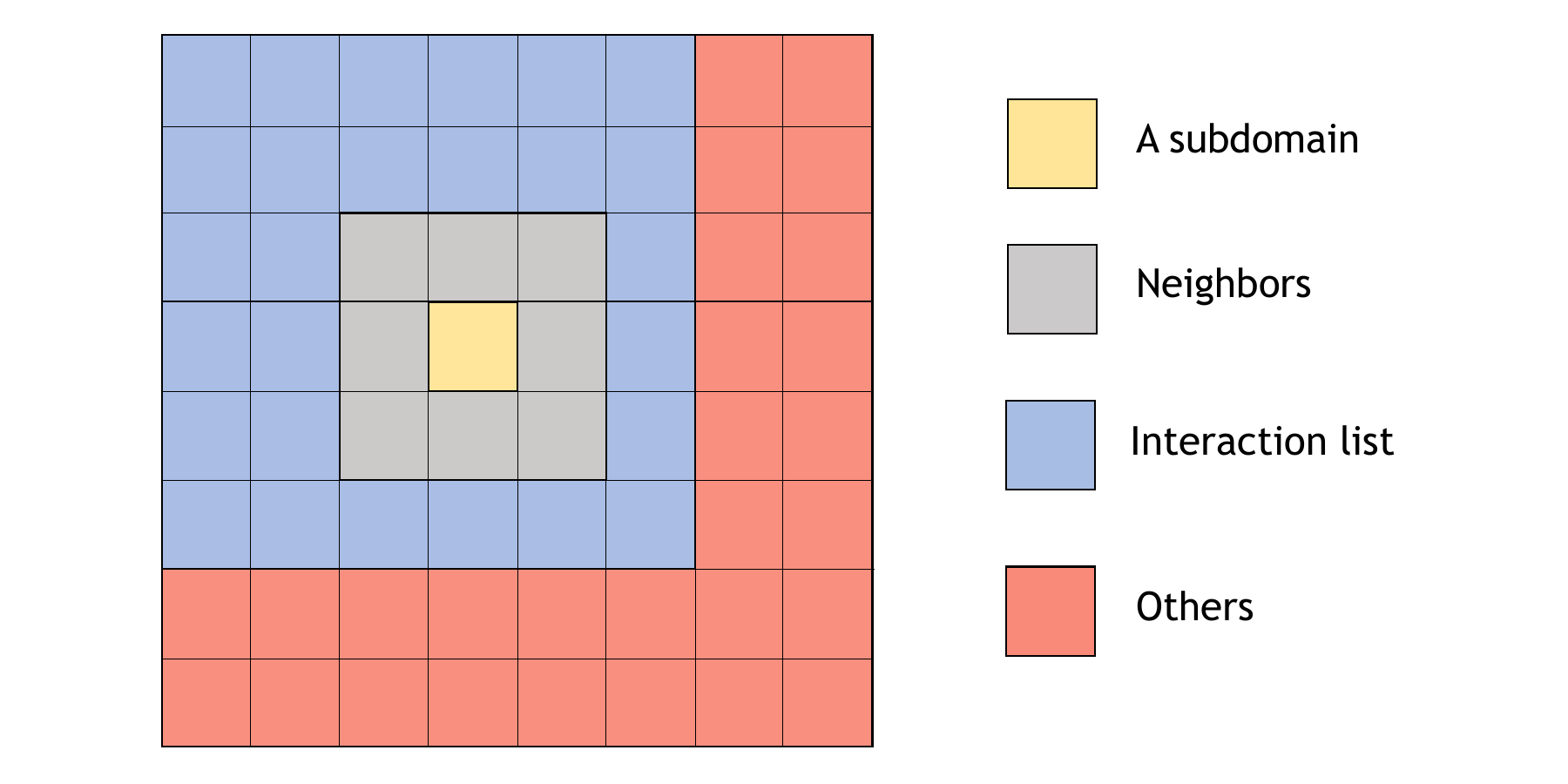}}
\caption{(a) hierarchical decomposition of the problem domain and the associated tree structure. The tree translation operators (M2M, M2L and L2L) are introduced in \cref{sec:fmm}. (b) a subdomain $\alpha$ at the fourth level in the hierarchy, its neighbors $\mathcal{N}(\alpha)$ and its interaction list $\mathcal{I}(\alpha)$.}
\label{fig:tree}
\end{center}
\end{figure}

As with other multilevel methods, our approach is based on a hierarchical decomposition of the problem domain to achieve linear complexity. Specifically, a cubical domain is divided into eight subdomains through binary partitioning along every coordinate. Then, every subdomain is divided recursively until the number of data points in every subdomain is less than a prescribed constant. This hierarchical domain decomposition is naturally associated with an octree tree data structure, where the root stands for the entire domain, and the other nodes stand for subdomains at different levels in the hierarchy; see \cref{fig:tree}(a). For the rest of the paper, we use the terms subdomain and tree node interchangeably.

{Note the hierarchical decomposition is generally non-uniform ($\mathcal{T}$ is adaptive). But we choose to use a uniform-tree data structure in \texttt{PBBFMM3D} for the ease of parallel implementation, which is reasonably efficient as long as the point distribution is not extremely irregular. With a uniform tree, some leaf nodes may end up having few data points. In \texttt{PBBFMM3D}, empty nodes are skipped in the algorithm, and the overhead of using a uniform tree compared with an adaptive one is from processing nodes that have only a few points.}


Given the tree $\mathcal{T}$, the \emph{parent} $\mathcal{P}(\alpha)$ and the \emph{children} $\mathcal{C}(\alpha)$ of a node $\alpha$ are naturally defined. We also define the neighbors $\mathcal{N}(\alpha)$ and the interaction list $\mathcal{I}(\alpha)$ of a node $\alpha$ as below; see \cref{fig:tree}(b) for an pictorial illustration.

\begin{definition}[Neighbors and interaction list]
Given a hierarchical tree structure, 
\begin{itemize}
    \item $\mathcal{N}(\alpha)$: the adjacent nodes (subdomains) of $\alpha$ at the same level in the hierarchy including $\alpha$ itself. The number of neighbors is generally $3^d$, where $d$ is the spatial dimension.
    \item $\mathcal{I}(\alpha) = \mathcal{C}(\mathcal{N}(\mathcal{P}(\alpha)))  / \mathcal{N}(\alpha)$: the non-adjacent nodes (subdomains) at the same level in the hierarchy whose parents are neighbors of $\mathcal{P}(\alpha)$. The number of nodes in the interaction list is generally $6^d - 3^d$, where $d$ is the spatial dimension.
\end{itemize}
\end{definition}


\subsection{Separation of variables (low-rank approximation)}
The key idea of our algorithm is to approximate the kernel function $\mathcal{K}(\vecx, \vecy)$ through polynomial interpolation when the target point $\vecx$ and the source point $\vecy$ are distant. In this section, we focus on the situation where $\vecx$ and $\vecy$ are inside two non-adjacent subdomains, respectively. Technically speaking, $\vecx$ and $\vecy$ are \emph{well-separated}. For the following discussion, we need a set of $p$ interpolation nodes $\mathcal{S}=\{x^*_1,x^*_2, \ldots,x^*_p\}$ on the real line. Then we can form interpolation nodes in 3D with tensor products:
\[
\mathcal{S}\otimes\mathcal{S}\otimes\mathcal{S} = \left\{\vecx^*_i=(x^*_{i_1},x^*_{i_2},x^*_{i_3}), i_1,i_2,i_3=1,2,\ldots,p \right\},
\]
where $i$ and $(i_1,i_2,i_3)$ are 1D-index and 3D-index of the $p\times p\times p$ grid, respectively, e.g., $i_1 = i \,\%\, p$, $i_2 = i \mod p$, and $i_3 = i \mod p^2$.

\begin{definition}[Lagrange basis polynomials]
The $p$-th order Lagrange basis polynomials in 1D are 
\[
\ell_p(x^*_i,x) = \Pi_{1\le k \le p, k\not=i} \frac{x-x^*_k}{x^*_i-x^*_k},
\quad i=1,2,\ldots,p.
\]
The $p$-th order Lagrange basis polynomials in 3D are tensor products of the $p$-th order Lagrange basis polynomials in 1D:
\[
L_p(\vecx^*_i, \vecx) = \ell_p(x^*_{i_1}, x_1) \ell_p(x^*_{i_2}, x_2) \ell_p(x^*_{i_3}, x_3), \quad i=1,2,\ldots,p^3,
\]
where $\vecx = (x_1, x_2, x_3) \in \mathbb{R}^3$, and  $\vecx^*_i=(x^*_{i_1},x^*_{i_2},x^*_{i_3}) \in \mathcal{S}\otimes\mathcal{S}\otimes\mathcal{S}$.
\end{definition}

\begin{definition}[$p$-th order polynomial interpolant of $\mathcal{K}(\vecx, \vecy)$]
Denote $\hat{\mathcal{K}}(\vecx, \vecy)$ as the $p$-th order polynomial interpolant of $\mathcal{K}(\vecx, \vecy)$, which is obtained through Lagrange interpolation on $\vecx$ and $\vecy$, respectively ($\vecx$ and $\vecy$ are well-separated):
\begin{align}
\label{eq:low-rank-form}
    \mathcal{K}(\vecx, \vecy) 
    \approx &\sum_{i=1}^{p^3} \mathcal{K}(\vecx^*_i, \vecy) L_p(\vecx^*_i,\vecx) \notag \\ 
    \approx &\sum_{i=1}^{p^3} \sum_{j=1}^{p^3} \mathcal{K}(\vecx^*_i, \vecy^*_j) L_p(\vecx^*_i,\vecx)  L_p(\vecy^*_j,\vecy) = \hat{\mathcal{K}}(\vecx, \vecy),
\end{align}
where $\vecx^*_i, \vecy^*_j \in \mathcal{S}\otimes\mathcal{S}\otimes\mathcal{S}$.
\end{definition}

Observe that constructing the approximation $\hat{\mathcal{K}}(\vecx, \vecy)$ requires only evaluations of the kernel function $\mathcal{K}(\vecx^*_i, \vecy^*_j)$.

The \texttt{PBBFMM3D} code offers two options for interpolation nodes, namely, Chebyshev nodes 
\[
\mathcal{S}=\left\{ \cos \left(\frac{2k-1}{2p} \pi \right), k=1,2,\ldots,p \right\}
\]
and equally spaced nodes
\[
\mathcal{S}=\left\{ \frac{k-1}{p-1} , k=1,2,\ldots,p \right\}.
\]
With the Chebyshev nodes, the approximation in \cref{eq:low-rank-form} is nearly optimal among polynomials of the same order. More importantly, the error decays as $\O(\rho^{-p})$ if the kernel function $K(\vecx, \vecy)$ is analytic and bounded in the ``Bernstein ellipse'' of foci 1 and -1 with semimajor and semiminor axis lengths summing to $\rho$~\cite{trefethen2013approximation}.
With equally spaced nodes, the matrix $\mathcal{K}(\vecx^*_i, \vecy^*_j)$ in \cref{eq:low-rank-form} is a block-Toeplitz-Toeplitz-block matrix, which has a reduced memory footprint and can be applied in $\O(p^3 \log(p))$ time using the FFT~\cite{chen2018fast}. Note the Lagrange interpolation of high degree over equally spaced nodes does not always converge, even for smooth functions, which is known as Runge’s phenomenon. In practice, we find low-order approximations sufficiently accurate in many applications. The scheme of Lagrange interpolation on uniform nodes can be stabilized by fitting a polynomial of degree $d < 2 \sqrt{p}$ using least-squares.

\subsection{Black-box FMM algorithm} \label{sec:fmm}

To evaluate \cref{eq:summation}, our algorithm has the following four stages.
\begin{enumerate}
\item 
\emph{Upward pass.} A post-order traversal of $\mathcal{T}$ is performed to compute the ``multipole coefficients'' of every subdomain, which encodes information of the source points and their weights contained in the subdomain. For every leaf node $\alpha$ in $\mathcal{T}$, the \emph{particle-to-moment} (P2M) translation is executed:
\[
\bm{M}^{\alpha}_i = \sum_{\vecy_j \in \alpha} L_p(\vecy^*_i, \vecy_j) \,\, \sigma_j, \quad i=1,2,\ldots,p^3,
\]
where $\vecy_j$ and $\vecy^*_i$ are the source points and interpolation nodes in $\alpha$, respectively.
For every non-leaf node $\alpha$ in $\mathcal{T}$, the \emph{moment-to-moment} (M2M) translation is executed:
\[
\bm{M}^{\alpha}_i = \sum_{\beta \in \mathcal{C}(\alpha)} \sum_{j=1}^{p^3} L_p(\vecy^*_{i}, \vecy^{*}_j) \,\, \bm{M}^{\beta}_j, \quad i=1,2,\ldots,p^3,
\]
where $\vecy^*_{i}$ and $\vecy^{*}_j$ are the interpolation nodes in $\alpha$ and $\beta$, respectively.
\item 
\emph{Far-field interaction.} The ``local coefficients'' of every node is computed using the ``multipole coefficients'' of its interaction list. For every node $\alpha$ in $\mathcal{T}$, the \emph{moment-to-local} (M2L) translation is executed:
\[
\bm{F}^{\alpha}_i = \sum_{\beta \in \mathcal{I}(\alpha)} \sum_{j=1}^{p^3} \mathcal{K}(\vecx^*_i, \vecy^*_j) \,\, \bm{M}^{\beta}_j,\quad i=1,2,\ldots,p^3,
\]
where $\vecx^*_{i}$ and $\vecy^{*}_j$ are the interpolation nodes in $\alpha$ and $\beta$, respectively.
\item 
\emph{Downward pass.} A pre-order traversal of $\mathcal{T}$ is performed to ``accumulate'' the ``local coefficients'' at leaf nodes. This is effectively a ``transpose'' of the upward pass. For every node $\alpha$ in $\mathcal{T}$, the \emph{local-to-local} (L2L) translation is executed:
\[
\bm{F}^{\alpha}_i \mathrel{+}= \sum_{j=1}^{p^3} \bm{F}^{\beta}_j \,\, L_p(\vecx^*_{j}, \vecx^*_{i}), \quad i=1,2,\ldots,p^3,
\]
where $\beta = \mathcal{P}(\alpha)$, $\vecx^*_{i}$ and $\vecx^*_{j}$ are the interpolation nodes in $\alpha$ and $\beta$, respectively.
For every leaf node $\alpha$ in $\mathcal{T}$, the \emph{local-to-particle} (L2P) translation is also executed:
\[
\phi_i = \sum_{j=1}^{p^3} \bm{F}^{\alpha}_j \,\, L_p(\vecx^*_{j}, \vecx_i) ,  
\]
where $\vecx_i$ and $\vecx^*_{j}$ are the target points and interpolation nodes in $\alpha$, respectively.
\item 
\emph{Near-field interaction.} The contribution from neighbors is evaluated exactly. For every leaf node $\alpha$, the \emph{particle-to-particle} (P2P) translation is executed:
\[
\phi_i \mathrel{+}= \sum_{\beta \in \mathcal{N}(\alpha)} \sum_{\vecy_j \in \beta} \mathcal{K}(\vecx_i, \vecy_j) \,\, \sigma_j,
\]
where $\vecx_i$ and $\vecy_j$ are the target and the source points in $\alpha$ and $\beta$, respectively.
\end{enumerate}

\begin{figure}[htbp]
\begin{center}
\includegraphics[trim=80 70 20 60,clip,width=15cm]{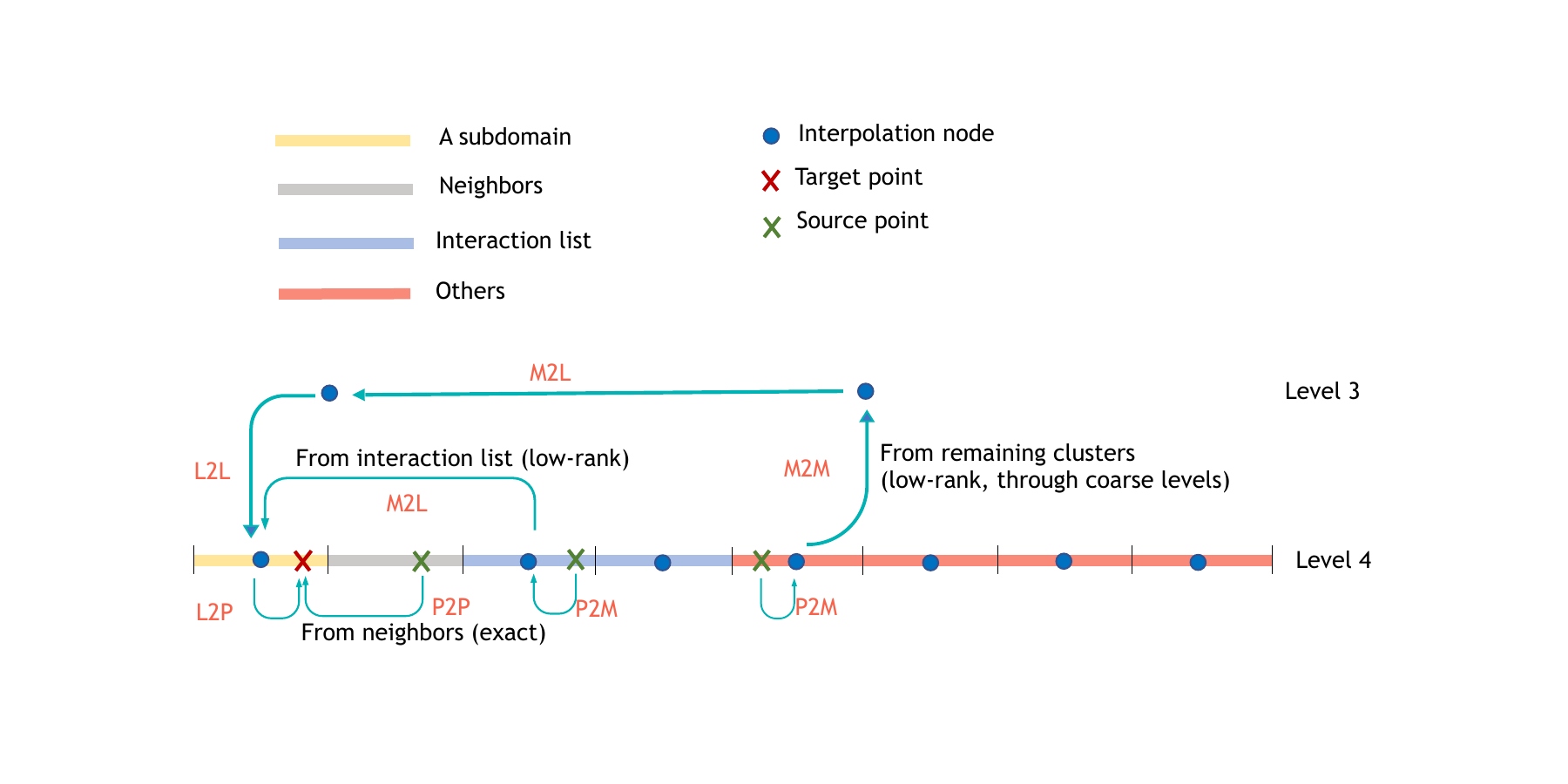}
\caption{A one-dimensional FMM example with one target point and three source points. The calculation of \cref{eq:summation} is divided into three parts: (1) P2P translation from the neighbor (grey), (2) P2M$\rightarrow$M2L$\rightarrow$L2P translations from the interaction list (blue), and (3) P2M$\rightarrow$M2M$\rightarrow$M2L$\rightarrow$L2L$\rightarrow$L2P translations from the remaining leaf nodes (red).
}
\label{fig:interactions}
\end{center}
\end{figure}

To summarize, the contribution from neighbors is calculated exactly; the contribution from the interaction list is approximated using the polynomial interpolant in \cref{eq:low-rank-form}; and the contribution from the remaining leaf nodes are approximated through coarser levels in the tree. A diagram illustrating the above process is shown in \cref{fig:interactions}. {Although our algorithm is presented using scalar operations for ease of illustration, it is implemented using BLAS2 and BLAS3 subroutines}.

Below, we present a theorem for the computational cost and memory footprint of the above algorithm. Our primary focus is showing the complexity in terms of the number of points $N$, and thus we present the computational cost of one M2L translation as $\O(p^6)$, the cost of a naive matrix-vector product. For the same reason, we use an $\O(p^6)$ estimate for the storage of every M2L translation operator. Given the M2L translation is a bottleneck in the FMM, we implement acceleration techniques in PBBFMM3D, which is discussed in \cref{sec:optimize}.

\begin{theorem}[Computational cost and memory footprint]
The computational cost and memory footprint of the black-box FMM algorithm are both $\O(N)$.
\end{theorem}

Suppose every leaf node has at most $n_0$ points (typically $64 \sim 128$). The number of leaf nodes and the number of tree nodes are both $\O(N/n_0)$. In the above algorithm, every node requires constant amount of work (independent of $N$) at every stage:
\begin{enumerate}
    \item \emph{Upward pass.} P2M: $\O(p^3 \, n_0)$ work for every leaf node. M2M: $\O(8 \, p^4)$ work for every non-leaf node, where every non-leaf node has at most 8 children in 3D. Note the M2M is a 3D tensor-vector multiplication.
    \item \emph{Far-field interaction.} M2L: $\O(189 \, p^6)$ work for every node, where the interaction list has at most $6^3 - 3^3 = 189$ nodes in 3D.
    \item \emph{Downward pass.} L2L: $\O(p^4)$ work for every node. L2P: $\O(p^3 \, n_0)$ work for every leaf node. Note the L2L is a 3D tensor(transpose of the tensor in M2M)-vector multiplication.
    \item \emph{Near-field interaction.} P2P: $\O(27 \, n_0^2)$ work for every leaf node, where every node generally has at most $3^3=27$ neighbors in 3D. 
\end{enumerate}
Therefore, the computational complexity of the entire algorithm is $\O(N)$. 

Regarding the memory footprint, every tree node stores $p^3$ ``multipole coefficients'' and $p^3$ ``local coefficients'', which sums up to $\O(Np^3/n_0)$. Since the kernel function is translational invariant, we need to precompute only $7^3-3^3=316$ M2L translation operators at every level for the $\log(N)$-level tree structure, which is $\O(316 p^6 \log(N))$ memory in total. Therefore, the memory footprint is $\O(2p^3N + 316 p^6 \log(N))$.


\subsection{Parallel algorithm}
In this section, we analyze the parallelism in each of the four stages in the FMM algorithm, and we have implemented a parallel algorithm using the \verb|OpenMP| API for shared-memory machines.
\begin{enumerate}
    \item \emph{Upward pass.} As stated earlier, the upward pass is a post-order traversal of $\mathcal{T}$, so a parallel post-order tree traversal using \verb|OpenMP| tasks is implemented in \texttt{PBBFMM3D}.
    \item \emph{Far-field interaction.} The M2L translations are independent for all nodes. In addition, the translation typically requires the same amount of work for every node. Therefore, the \verb|OpenMP| ``parallel for'' directive is employed on the loop over all nodes for M2L translations in \texttt{PBBFMM3D}.
    \item \emph{Downward pass.} As stated earlier, the downward pass is a pre-order traversal of $\mathcal{T}$, so a parallel pre-order tree traversal using \verb|OpenMP| tasks is implemented in \texttt{PBBFMM3D}.
    \item \emph{Near-field interaction.} The P2P translations are independent for all leaf nodes. However, the translations between pairs of neighbors are work-heterogeneous due to the non-uniform distribution of the target and the source points. In \texttt{PBBFMM3D}, the \verb|OpenMP| ``parallel for'' directive is employed on the loop over all leaf nodes for P2P translations, and different scheduling policies can used based on any prior knowledge of the point distribution.
\end{enumerate}

\cref{alg:fmm} shows the pseudocode of the above parallel algorithm. A common alternative to the parallel tree traversal using \verb|OpenMP| task is a level-by-level traversal with the \verb|OpenMP| ``parallel for'' directive on the loop over all nodes at the same level. Although the P2M and the L2P translations are work-heterogeneous, the efficiency of the alternative approach may still be reasonable since the upward pass and the downward pass are usually not the bottleneck.

\begin{algorithm}[hbtp]
\caption{Black-box FMM algorithm, where subroutines are shown in \cref{alg:subroutine} (ignoring the ``pragma'' lines leads to the serial algorithm)}
\label{alg:fmm}
\begin{algorithmic}[1]
\State \emph{\# pragma omp parallel}
\State \emph{\# pragma omp single}
\State \Call{Upward\_Pass}{root of $\mathcal{T}$}
\State \Call{Far\_field\_interaction()}{}
\State \emph{\# pragma omp parallel}
\State \emph{\# pragma omp single}
\State \Call{Downward\_Pass}{root of $\mathcal{T}$}
\State \Call{Near\_field\_interaction()}{}
\end{algorithmic}
\end{algorithm}

\begin{algorithm}[hbtp]
\caption{Subroutines in the FMM algorithm}
\label{alg:subroutine}
\begin{algorithmic}[1]
\Function{Upward\_Pass}{node $\alpha$}
\ForAll{node $\beta \in \mathcal{C}(\alpha)$}
\State \emph{\# pragma omp task}
\State \Call{Upward\_Pass}{$\beta$}
\EndFor
\State \emph{\# pragma omp taskwait}
\If{$\alpha$ is a leaf node} 
\State P2M translation
\Else 
\State M2M translation
\EndIf
\EndFunction
\State
\Function{Far\_field\_interaction()}{}
\State \emph{\# pragma omp parallel for}
\ForAll{node $\alpha$ in $\mathcal{T}$} 
\State M2L translation
\EndFor
\EndFunction
\State
\Function{Downward\_Pass}{node $\alpha$}
\If{$\alpha$ is a leaf node} 
\State L2P translation
\Else 
\State L2L translation
\EndIf
\ForAll{node $\beta \in \mathcal{C}(\alpha)$}
\State \emph{\# pragma omp task}
\State \Call{Downward\_Pass}{$\beta$}
\EndFor
\EndFunction
\State
\Function{Near\_field\_interaction()}{}
\State \emph{\# pragma omp parallel for reduction(+: $\bm{\phi}$)}
\ForAll{leaf node $\alpha$ in $\mathcal{T}$} 
\State P2P translation
\EndFor
\EndFunction
\end{algorithmic}
\end{algorithm}

Here our parallel algorithm focuses on parallelizing each stage of the FMM algorithm, and haven't exploit the concurrency across different stages. In principle, the near-field interaction stage does not depend on the others except for updating the results. So the work-heterogeneous P2P translations can be prioritized. Furthermore, notice the M2L translation of a node can happen as soon as its ``multipole coefficients'' have been computed. However, implementing these ideas efficiently typically requires a task-based runtime system, for which we refer interested readers to~\cite{agullo2014task,agullo2016task}.

\subsection{Acceleration techniques} \label{sec:optimize}
Since the far-field interaction and the near-field interaction usually dominate the entire computation, we introduce the acceleration techniques used in \texttt{PBBFMM3D}.

\paragraph{Homogeneous kernel} 
A kernel function is homogeneous if
\[
\mathcal{K}(\alpha \vecx, \alpha \vecy) = \alpha^m \mathcal{K}(\vecx, \vecy),
\]
for $\alpha \not= 0$ and $m$ is typically an integer. For example, $\mathcal{K}(\vecx, \vecy) = 1/\|\vecx - \vecy\|$ is a homogeneous kernel function of degree -1. Since the interpolation grids are fixed relative to the problem domain, the M2L translation operators of different levels differ only by a scaling constant. Hence, we store these operators for only the leaf level.

\paragraph{Symmetry and skew-symmetry} 
A kernel function is symmetric if
\[
\mathcal{K}( \vecx,  \vecy) = \mathcal{K}(\vecy, \vecx),
\]
and is skew-symmetric if
\[
\mathcal{K}( \vecx,  \vecy) = - \mathcal{K}(\vecy, \vecx).
\] 
In many applications, the source $\{\vecx_i\}_{i=1}^N$ and the target $\{\vecy_i\}_{i=1}^N$ are the same set of points. Therefore, the kernel matrix $\bm{K}$ becomes symmetric or skew-symmetric if the kernel function is such. This implies only one of the two P2P/M2L translation operators needs be stored between a pair of tree nodes that are either neighbors or in each other's interaction list.

\paragraph{Fast M2L translation}
Recall the definition of an M2L translation operator $\mathcal{K}(\vecx^*, \vecy^*)$ in \cref{eq:low-rank-form}, where $\vecx^*$ and $\vecy^*$ are interpolation grids in a pair of tree nodes that are in each other's interaction list. In \texttt{PBBFMM3D}, there are two options for the interpolation grids: Chebyshev nodes or equally spaced nodes. With Chebyshev nodes, an SVD-based compression of the M2L operator is employed following the approach in~\cite{fong2009black} since the translation operator is observed to be numerically low rank. 
With equally spaced nodes, the M2L operator is a block-Toeplitz-Toeplitz-block matrix~\cite{chen2018fast}, where the $p^3$-by-$p^3$ matrix has only $(2p-1)^3$ unique entries and can be applied to a vector in $\O(p^3 \log(p))$ time using the FFT.

\paragraph{Multiple right-hand-sides} 
In some applications, \cref{eq:summation} needs to be evaluated with multiple weight vectors associated with the same set of source points. In \texttt{PBBFMM3D}, all weight vectors are grouped into a matrix as the input of the FMM algorithm, which allows using cache-friendly BLAS3 operations in our algorithm.

\section{Software description.}
\label{sec:software_description}

In this section, we briefly discuss the software architecture of \texttt{PBBFMM3D} focusing on the black-box feature of the algorithm and the \verb|C++| and \verb|Python| interfaces. More details can be found in the documentation at \url{https://github.com/ruoxi-wang/PBBFMM3D}. The code is written in \verb|C++| with the \verb|OpenMP| API, and requires some standard linear algebra libraries including the BLAS\footnote{\url{http://www.netlib.org/blas/}}, the LAPACK\footnote{\url{http://www.netlib.org/lapack/}} and the FFTW3 library\footnote{\url{http://www.fftw.org/}}. The Boost \verb|Python| Libraries\footnote{\url{https://www.boost.org/doc/libs/1_70_0/libs/python/doc/html/index.html}} is required for using the \verb|Python| interface.

\begin{figure}[htbp]
\centering
\includegraphics[trim=70 5 15 20, clip, width=1.0\textwidth]{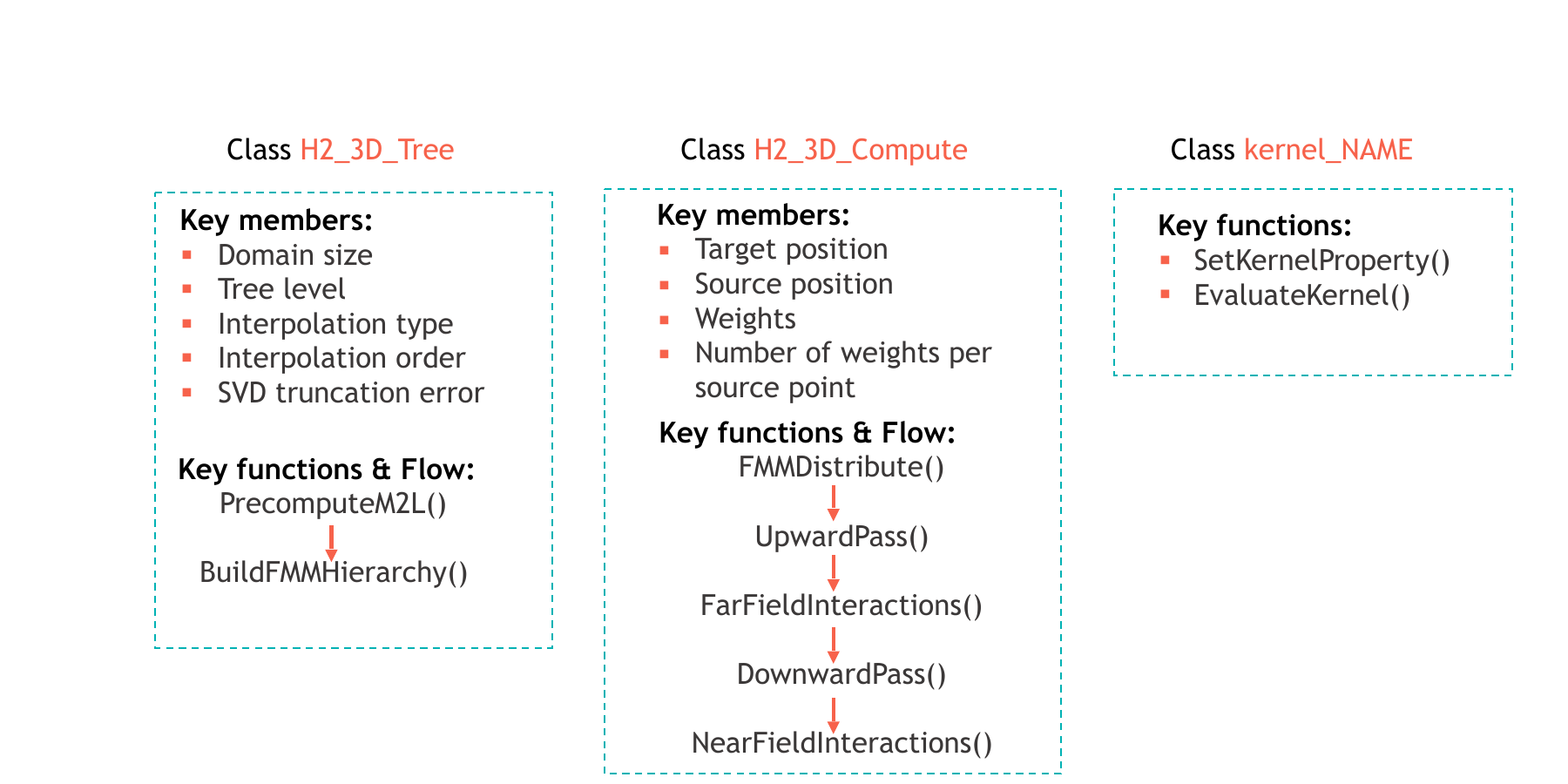}
\caption{Three main classes in \texttt{PBBFMM3D}. The \texttt{H2\_3D\_Tree} class stores information regarding the hierarchical partitioning of the problem domain. The \texttt{H2\_3D\_Compute} class implements the FMM algorithm, and the \texttt{kernel\_NAME} class describes the kernel function.}
\label{fig:architecture}
\end{figure}

The \texttt{PBBFMM3D} has the following three main classes as shown in \cref{fig:architecture}.
\begin{itemize}
    \item 
    Class \verb|H2_3D_Tree| sets parameters and creates the hierarchical partitioning of the problem domain. The parameters include (1) ``\texttt{Domain size}'': side length of a cubical problem domain, (2) ``\texttt{Tree level}'': the number of levels in the hierarchical partitioning, (3) ``\texttt{Interpolation type}'': Chebyshev nodes or equally spaced nodes, (4) ``\texttt{Interpolation order}'': the number of interpolation nodes used in \cref{eq:low-rank-form}, (5) ``\texttt{SVD truncation error}'': error from the compression of the M2L translation operators, which is by default the prescribed accuracy of the entire computation.
    The two key member functions are the following. Function \verb|PrecomputeM2L()| precomputes the M2L operators, and \verb|BuildFMMHierarchy()| creates the hierarchical partitioning and builds the corresponding data structure. 
    \item
    Class \verb|H2_3D_Compute| stores the information regarding the source and the target points including (1) ``\texttt{Target}'': position of target points $\{\vecx_i\}_{i=1}^N$, (2) ``\texttt{Source}'': position of source points $\{\vecy_i\}_{i=1}^N$, (3) ``\texttt{Weight}'': weights $\{\sigma_i\}_{i=1}^N$ associated with the source points, and (4) ``\texttt{Number of weights}'': number of weights associated with every source point.  
    The five key member functions include \verb|FMMDistribute()|, which assigns the source and the target points to leaf cells in the tree, and four functions correspond to the four translation stages described in \cref{sec:algorithm}.
    \item
    Class \verb|kernel_NAME| defines the kernel function. The member function \verb|SetKernelProperty()| sets the homogeneous and the symmetric properties of the kernel function. The member function \verb|EvaluateKernel()| takes two data points $\vecx$ and $\vecy$ and returns the value of $\mathcal{K}(\vecx, \vecy)$.
\end{itemize}

\paragraph{C++ \& Python interfaces}
Listing~\ref{lst:cpp} and~\ref{lst:python} show the core lines of a basic example using the C++ interface and the Python interface, respectively. The example evaluates \cref{eq:summation} with the standard Gaussian kernel. The first line creates an object of Class \verb|kernel_Gaussian|, which implements the standard Gaussian function. The class inherits from class \verb|H2_3D_Tree| and takes input parameters. The second line creates the hierarchical partitioning of the problem domain. The last line does the computation and stores results in the output variable.


\begin{lstlisting}[language=c++,caption={C++ interface. ``\texttt{tree}'' is an object of class \texttt{kernel\_Gaussian}, which implements the standard Gaussian kernel. It also stores input parameters for the hierarchical partitioning. ``\texttt{compute}'' takes the positions of the target points, the source points, the associated weight vector(s), and the number of weight vectors. It evaluates \cref{eq:summation} and outputs the results.},captionpos=b,label={lst:cpp}]
// define kernel and set parameters for hierarchical partitioning
kernel_Gaussian tree(domain, level, IP_type, IP_order, SVD_err);
// partition
tree.buildFMMTree();
// FMM
H2_3D_Compute<kernel_Gaussian> compute(tree, target, source, weight, num_weight, result);
\end{lstlisting}

\begin{lstlisting}[language=Python,caption={Python interface for the example in Listing~\ref{lst:cpp}.},captionpos=b,label={lst:python}]
from FMMTree import *
from FMMCompute import *
# read/set inputs and parameters
...
# define kernel and set parameters for hierarchical partitioning
tree = kernel_Gaussian(domain, level, IP_type, IP_order, SVD_err)
# partition
tree.buildFMMTree()
# FMM
Compute(tree, target, source, weight, num_weight, result)
\end{lstlisting}


\paragraph{Customized kernel}
While some commonly used kernel functions are already implemented in \texttt{PBBFMM3D}, defining a new kernel function is straightforward as shown in Listing~\ref{lst:myKernel}. It requires implementing only two methods as follows. The \verb|EvaluateKernel()| method takes a pair of source and target points and returns the function value, and the \verb|SetKernelProperty()| method tells the homogeneous degree of the kernel and whether it is symmetric (see \cref{sec:optimize}).

\begin{lstlisting}[language=c++,caption={Example of defining the $e^{-\|x-y\|}$ kernel. The function is symmetric but not homogeneous. Here, ``vector3'' is a structure of three floating point numbers representing coordinates in 3D.},captionpos=b,label={lst:myKernel}]
class myKernel: public H2_3D_Tree {
  public:
    myKernel(double domain, int level, int IP_type, int IP_order, double SVD_err):
      H2_3D_Tree(domain, level, IP_type, IP_order, SVD_err) {};
    virtual void SetKernelProperty() {
      homogen = 0;
      symmetry = 1;
      kernelType = "exponential";
    }
    virtual double EvaluateKernel(const vector3 &target, const vector3 &source) {
      vector3 diff;
      diff.x = source.x - target.x;
      diff.y = source.y - target.y;
      diff.z = source.z - target.z;
      double r = std::sqrt(diff.x*diff.x+diff.y*diff.y+diff.z*diff.z);
      return std::exp(-r);
    }
};
\end{lstlisting}

\section{Numerical Results}
\label{sec:examples}
In this section, we present numerical experiments with \texttt{PBBFMM3D} to show the accuracy, the sequential running time and the parallel scalability. A real-world application in geostatistics is also presented, where \texttt{PBBFMM3D} was used to speed up the calculations with a covariance matrix. We focus on two particular kernel functions here: ${1}/{\|\vecx-\vecy\|}$ and $\text{exp}(-\|\vecx-\vecy\|)$, where ${1}/{\|\vecx-\vecy\|}$, the Green's function for the Laplace equation in 3D, is frequently used in computational physics; and $\text{exp}(-\|\vecx-\vecy\|)$ is a popular choice as the covariance function for a Gaussian process.

All experiments were performed on a Linux server with 192 GB of RAM and the Intel Xeon Platinum 8280 (``Cascade Lake") with 56 cores on two sockets (28 cores/socket). The code was compiled with \texttt{GCC} 8.3.0, which implements version 4.5 of the \texttt{OpenMP} standard, and the code was linked with the Intel MKL library\footnote{https://software.intel.com/en-us/mkl}  version 2020.1.217 and the FFTW3 library version 3.3.8. 

\paragraph{Parameters and notations}
\begin{itemize}
    \item $N$: the number of source/target points, which are randomly generated using the uniform distribution in the unit cube.
    \item $p$: the order of interpolation, which determines the accuracy of our algorithm. Note the number of interpolation nodes is $p^3$ in \cref{eq:low-rank-form} as a tensor product of $p$ nodes in each dimension.
    \item \texttt{unif}: using equally spaced nodes in \cref{eq:low-rank-form}.
    \item \texttt{cheb}: using Chebyshev nodes in \cref{eq:low-rank-form}.
    \item \texttt{tree\_level}: the number of levels of the hierarchical partitioning of the problem domain.
    \item SVD accuracy: the compression accuracy of the M2L translation operator, which is chosen to be the same as the prescribed accuracy.
    \item $r$: notation of the Euclidean distance between a pair of source point $\vecx$ and a target point $\vecy$, i.e., $r=\|\vecx-\vecy\|_2$. 
\end{itemize}

\subsection{Accuracy \& sequential running time}
In this section, we focus on two parameters $N$ and $p$. First, we fix $N=10^4$ and increase $p$ to show the accuracy of \texttt{PBBFMM3D}. (We refer to~\cite{chen2018fast} for the precomputation time and the memory footprint.) Then, we fix $p=4$ and increase $N$ to show the sequential running time. 

\begin{figure}[htbp]
\centering
\subfigure[$1/r$]{\includegraphics[width=0.45\linewidth]{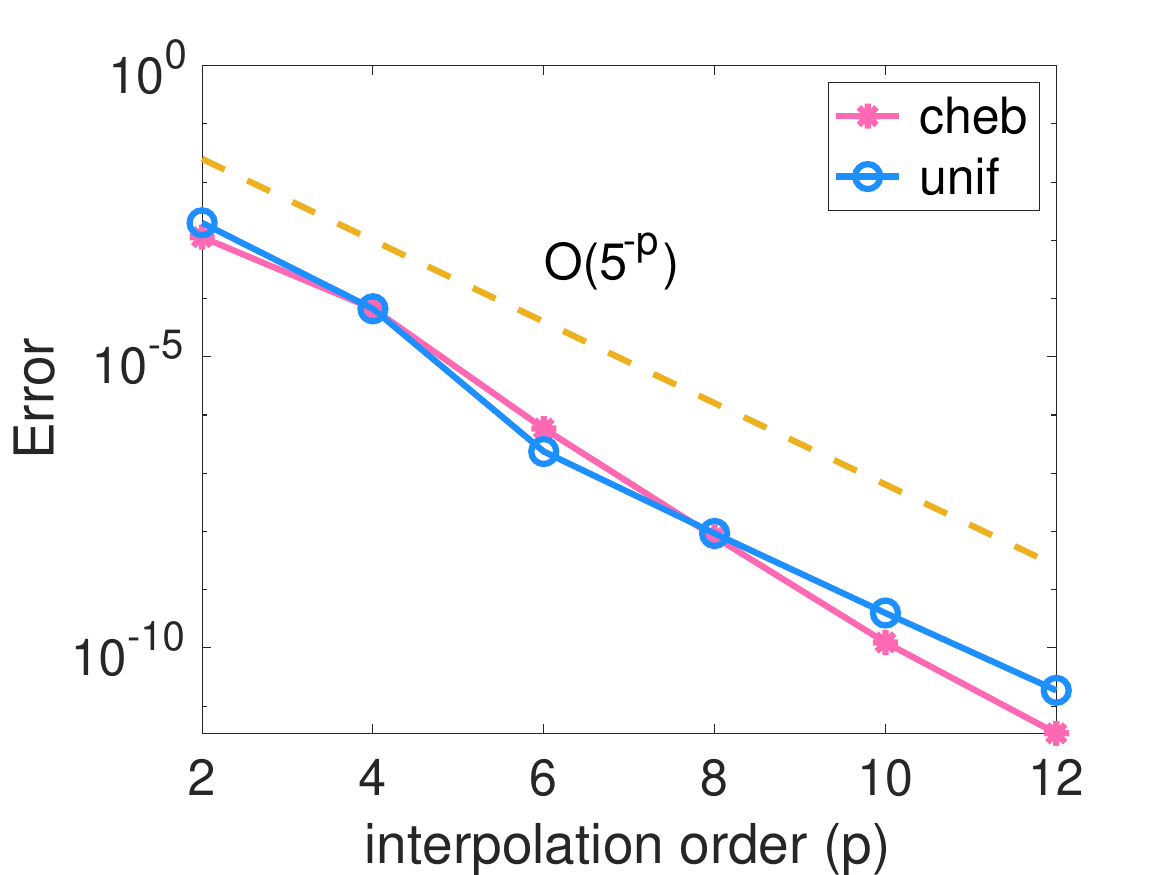}}
\subfigure[$\exp(-r)$]{\includegraphics[width=0.45\linewidth]{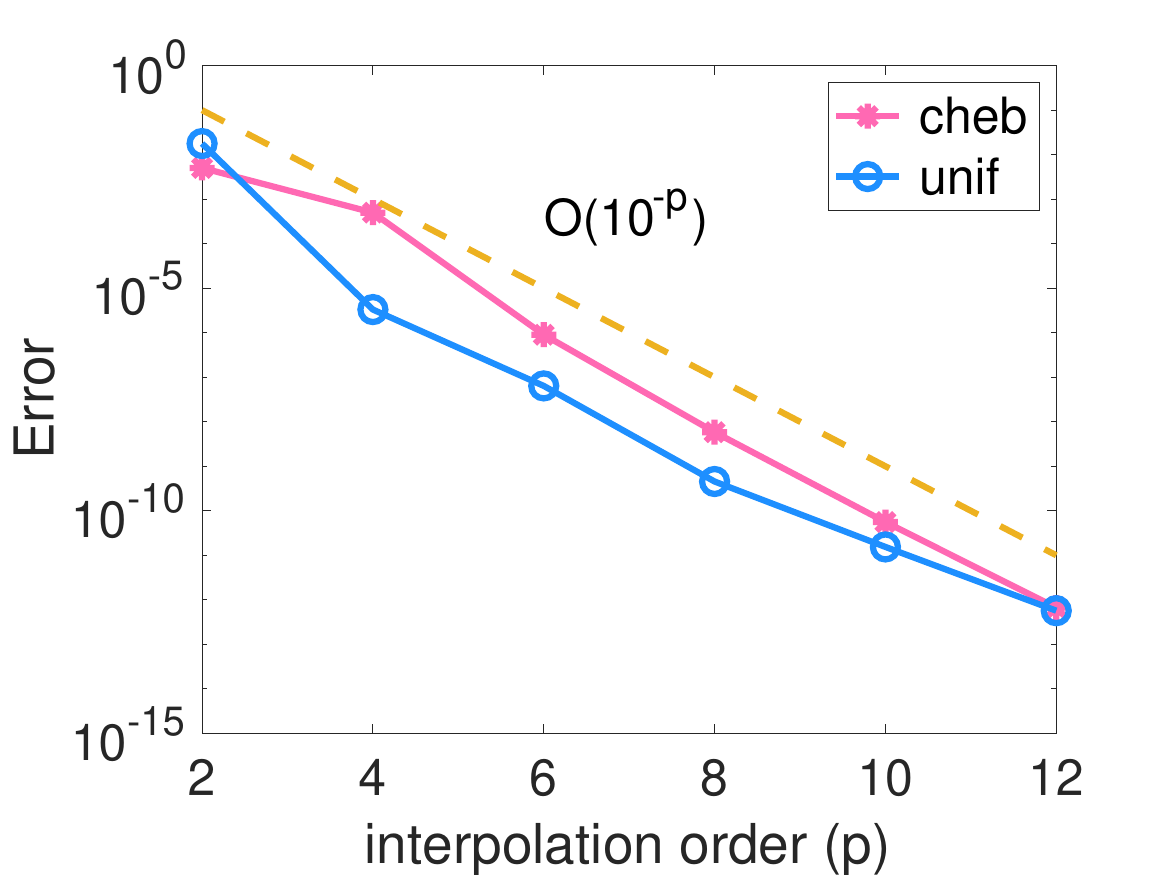}}
\caption{Relative error vs.\ interpolation order $p$, where $N=10^4$ and $\texttt{tree\_level}=5$. Both \texttt{unif} and \texttt{cheb} lead to exponential decay of errors.}
\label{fig:err_order}
\centering
\subfigure[$1/r$]{\includegraphics[width=0.45\linewidth]{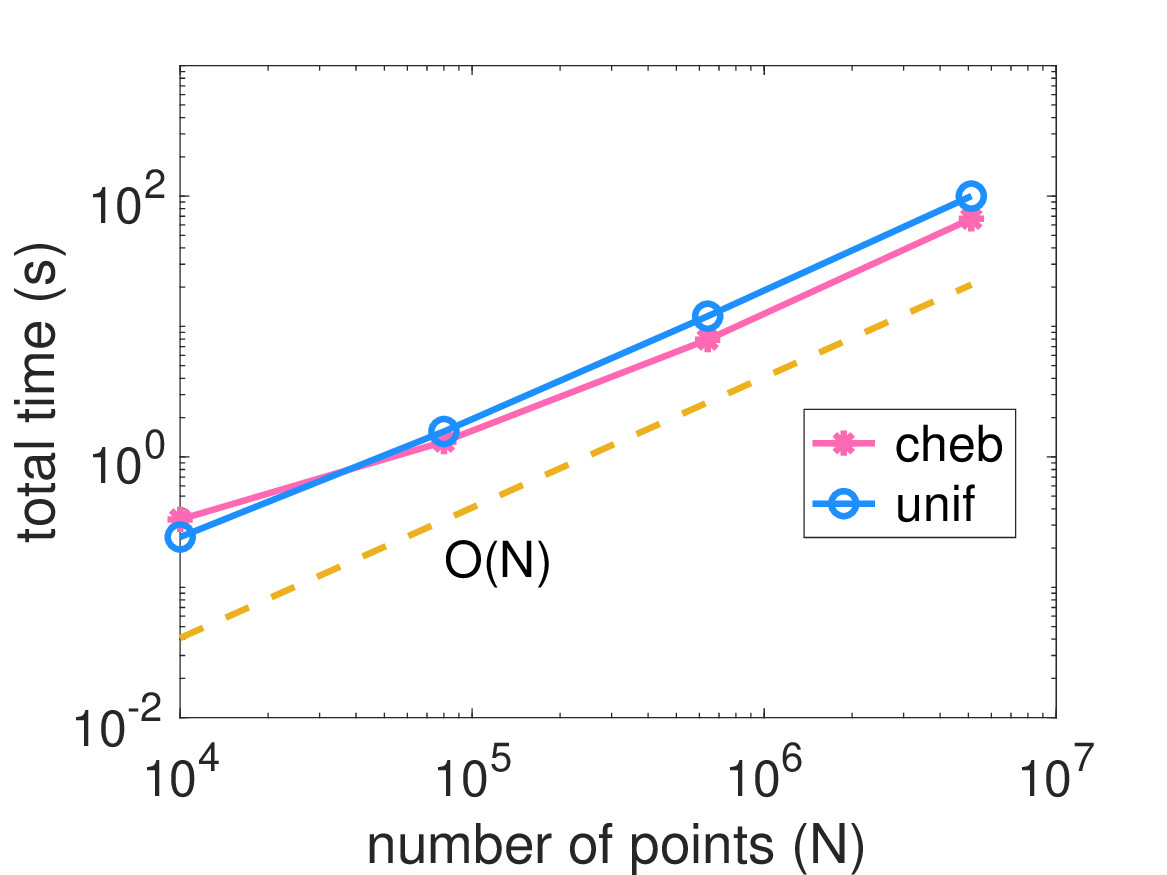}}
\subfigure[$\exp(-r)$]{\includegraphics[width=0.45\linewidth]{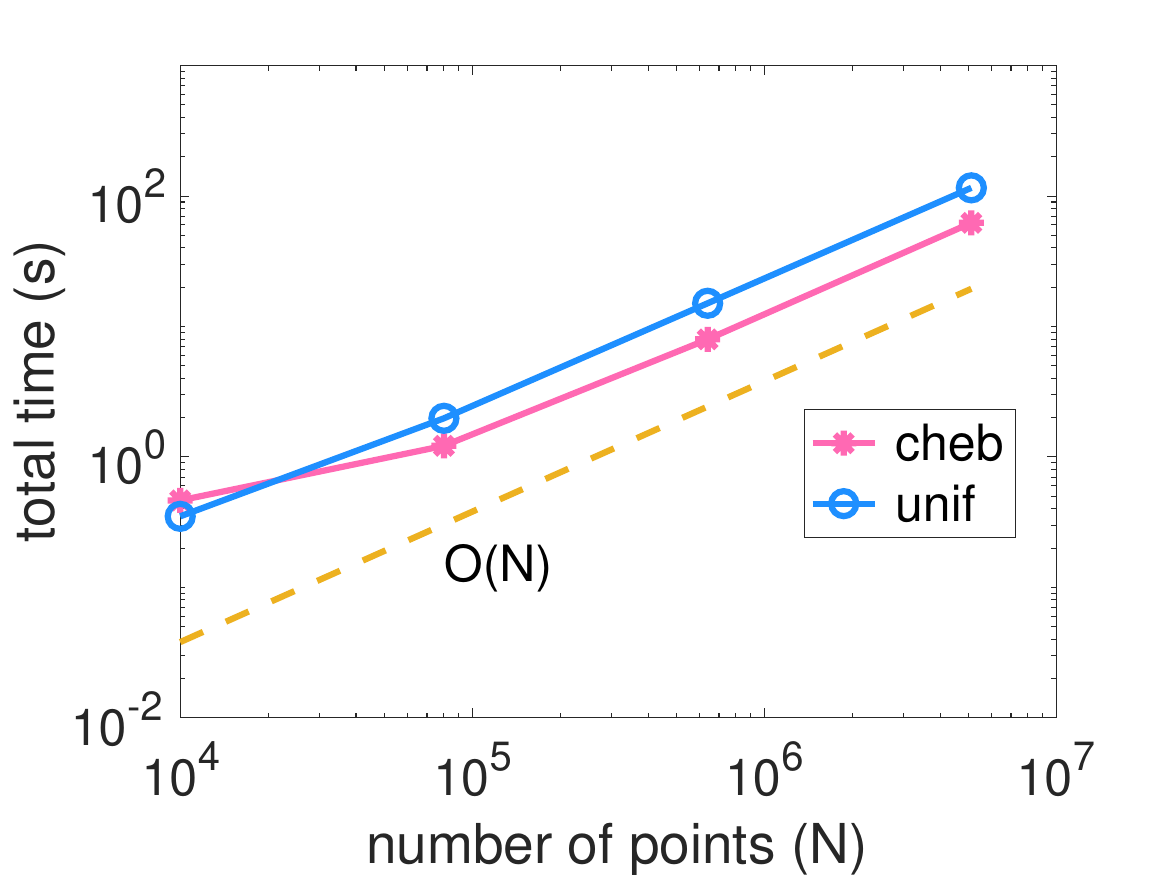}}
\caption{Sequential running time vs.\ number of points ($N = 10^4, 8\times 10^4, 8^2\times 10^4, \text{and } 8^3\times 10^4$). The interpolation order $p$ is fixed at 4.}
\label{fig:linear_scale}
\end{figure}

\cref{fig:err_order} shows the (relative) error of evaluating \cref{eq:summation} as a function of $p$, where the kernel functions are $1/r$ and $\text{exp}({-r})$, respectively. The error is defined as 
\[
\frac{\|\hat K \vecsigma - K\vecsigma\|_2}{\|K\vecsigma\|_2},
\]
where $\hat K$ stands for the approximation  constructed implicitly in \texttt{PBBFMM3D}. Since both kernel functions are analytic (away from the origin), the error of our interpolation using either Chebyshev nodes or uniform nodes decays exponentially.

\cref{fig:linear_scale} shows the scaling of the sequential running time with respect to the number of points $N$. We see that the time increases linearly as $N$ increases, as opposed to the quadratic increase of evaluating \cref{eq:summation} naively.

\subsection{Parallel scalability}\label{sec:result_parallel}
In this section, we present the parallel running time of \texttt{PBBFMM3D} on up to 32 cores. We focus on the kernel function $1/r$ and use Chebyshev nodes in \cref{eq:low-rank-form} with $p=4$. 
\cref{tab:parallel_scalability} reports the parallel running time. 
 {To show the parallel scalability, we chose the number of cores to be a power of 2. But that is not required  in  \texttt{PBBFMM3D}. The baseline (1 core) is the serial version used in \cite{fong2009black, chen2018fast}.} As the table shows, we obtained approximately $19\times$ speedup on 32 cores for $N = 8^4 \times 10^4$.

\begin{table}[htbp]
\caption{Parallel running time of \texttt{PBBFMM3D}.  The kernel function is $1/r$, and we used Chebyshev interpolation with $p=4$. We set the number of tree levels to be 5, 6 and 7 for the three increasing problem sizes, and the errors are $2.10\text{e-}{5}$, $2.08\text{e-}{5}$, and $2.10\text{e-}{5}$, respectively ({independent of the number of cores used}).}
\centering
\begin{tabular}{c c c c c c c}
\toprule
\multirow{2}{*}{$N$} & \multicolumn{6}{c}{\bf Time (seconds) } \\
&    1 core &  2 cores &  4 cores &  8 cores & 16 cores & 32 cores\\
\midrule
$8^2$e+4 & $5.74$e+0 & $3.32$e+0   &$1.83$e+0 & $9.93$e-1 & $5.84$e-1& $4.44$e-1\\
$8^3$e+4 & $4.72$e+1 & $2.58$e+1 & $1.40$e+1 & $7.87$e+0 &$4.40$e+0 & $3.49$e+0\\
$8^4$e+4 & $4.04$e+2 & $2.15$e+2 & $1.14$e+2& $6.37$e+1 &$3.72$e+1 & $2.13$e+1\\
\bottomrule
\end{tabular}
\label{tab:parallel_scalability}
\end{table}

\begin{figure}[htbp]
\centering
\subfigure[Strong scalability]{\includegraphics[width=0.45\linewidth]{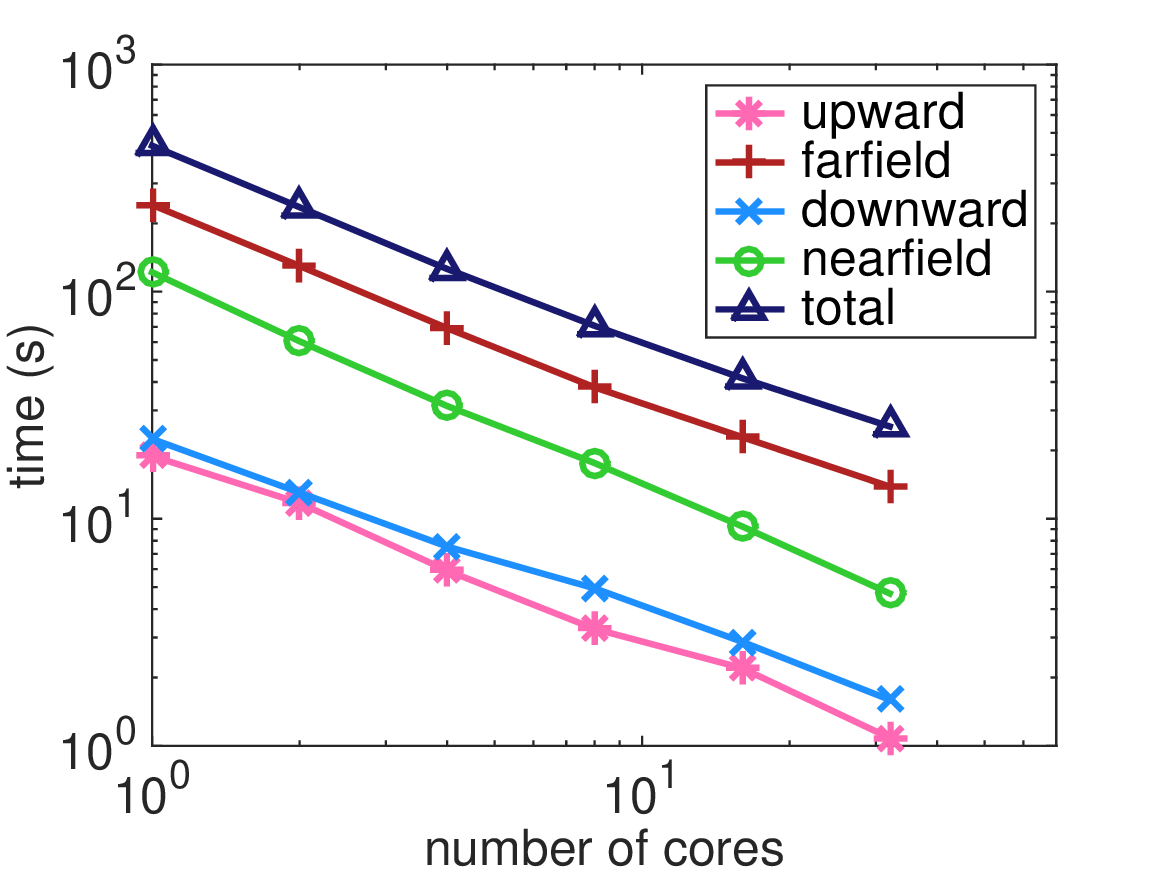}}
\subfigure[Weak scalability]{\includegraphics[width=0.53\linewidth]{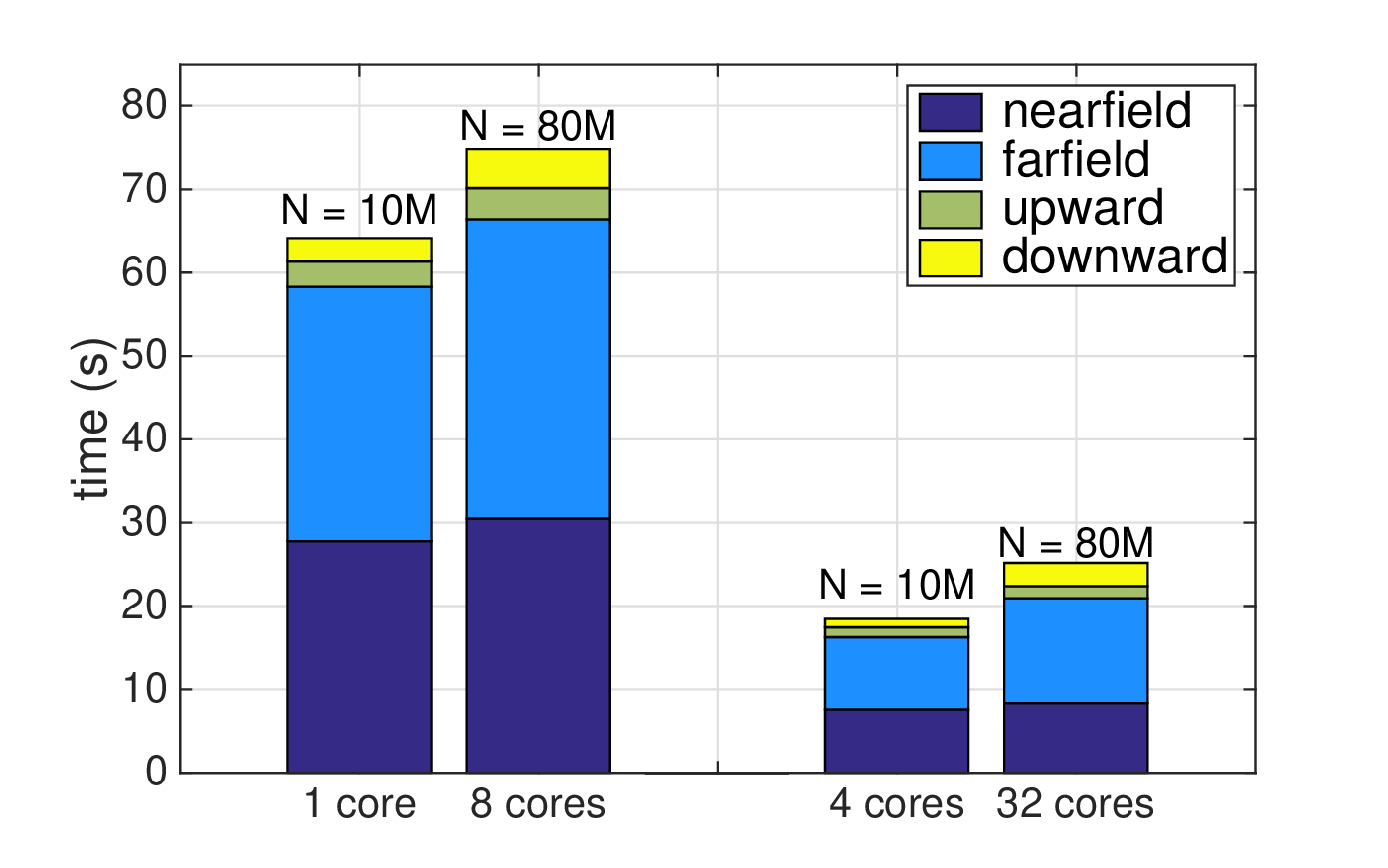}}
\caption{Parallel scalability and breakdown of the total time. The four stages in the algorithm are defined in \cref{sec:algorithm}. Strong scalability results correspond to a fixed problem size of $8^4 \times 10^4$ points. {The base cases for weak scalability are $2^{10} \times 10^4$ ($\sim 10$ million) points on 1 core and 4 cores, respectively.}}
\label{fig:parallel_scale}
\end{figure}

\cref{fig:parallel_scale}(a) shows the strong scalability of \texttt{PBBFMM3D}, i.e., the running time using a sequence of increasing number of cores for a fixed problem size. We provided a breakdown of the running time into the four stages: upward pass, far-field interaction, downward pass, and near-field interaction. We see that the running time for all stages nearly halved as the number of cores doubled.

\cref{fig:parallel_scale}(b) shows the weak scalability of \texttt{PBBFMM3D}, i.e., the running time for a fixed problem size per core. So we increased the number of particles proportionally to the number of cores. As the figure shows, the time spent on each stage in the FMM only increased by a small amount when the problem size increased by $8\times$ (the number of cores also increased by $8\times$). The ideal runtime would stay unchanged due to the linear complexity of \texttt{PBBFMM3D}. {Our implementation achieved an efficiency of 73\% to 86\% (serial time/parallel time).}

\subsection{Application in Gaussian Processes}

Gaussian random field (GRF) theory \cite{rasmussen2005gaussian,stein1999interpolation} has been widely used in interpolation and estimation of spatially correlated unknowns. For example, GRF methods can be used for estimating the permeability of the underground soil and rock, which is of critical interest to hydrogeologists and petroleum engineers \citep{kitanidis1997introduction,oliver2008inverse}. In GRF methods,  a covariance matrix is required as the prior information of the underlying unknown field. However, practical applications typically require large numbers of unknowns, so dimension reduction techniques such as the principle component analysis are required. 

To obtain the top-$k$ principle components, we employ \texttt{PBBFMM3D} to compute the truncated eigenvalue decomposition of the covariance matrix. In particular, we use a randomized method (Algorithm 5.3 in \cite{halko2011finding})  to calculate the top-$k$ eigenvalues and their associated eigenvectors, and the randomized algorithm requires evaluating \cref{eq:summation} $\O(k)$ times (same target and source points but with $\O(k)$ different weights). While the original method requires $\O(k N^2)$ operations, we accelerate the  method with \texttt{PBBFMM3D} and arrive at the optimal $\O(k N)$ complexity, where $N$ is the number of data points.

In the next experiment, we randomly generated data points in the unit cube and employed the kernel function $e^{-r}$, one-dimensional Mat\'{e}rn kernel with smoothness 1/2. The goal was to compute the top-100 eigen-pairs of the covariance matrix, where {120 matrix-vector products (packed into one matrix-matrix product) were evaluated twice in the randomized algorithm. In the original randomized method, matrix-matrix products are computed by calling the \texttt{dgemm}  subroutine in the Intel MKL library (with multi-threading on 32 cores).}
 In the \texttt{PBBFMM3D}-accelerated method, the number of   Chebyshev nodes was $p=4$, and  the number of tree levels  were 3, 4 and 5 for the three increasing problem sizes. We measured the error  $\|\Lambda_{FMM} - \Lambda_{ext}\|_2 / \|\Lambda_{ext}\|_2$, where $\Lambda_{ext}$ and $\Lambda_{FMM}$ are the  eigenvalues computed using the original method and the \texttt{PBBFMM3D}-accelerated method, respectively. The errors for $N=10^4$ and $N=8 \times 10^4$ are $1.6 \times 10^{-4}$ and $2.8 \times 10^{-5}$, respectively.

\begin{figure}
\begin{center}
\includegraphics[width=0.5\textwidth]{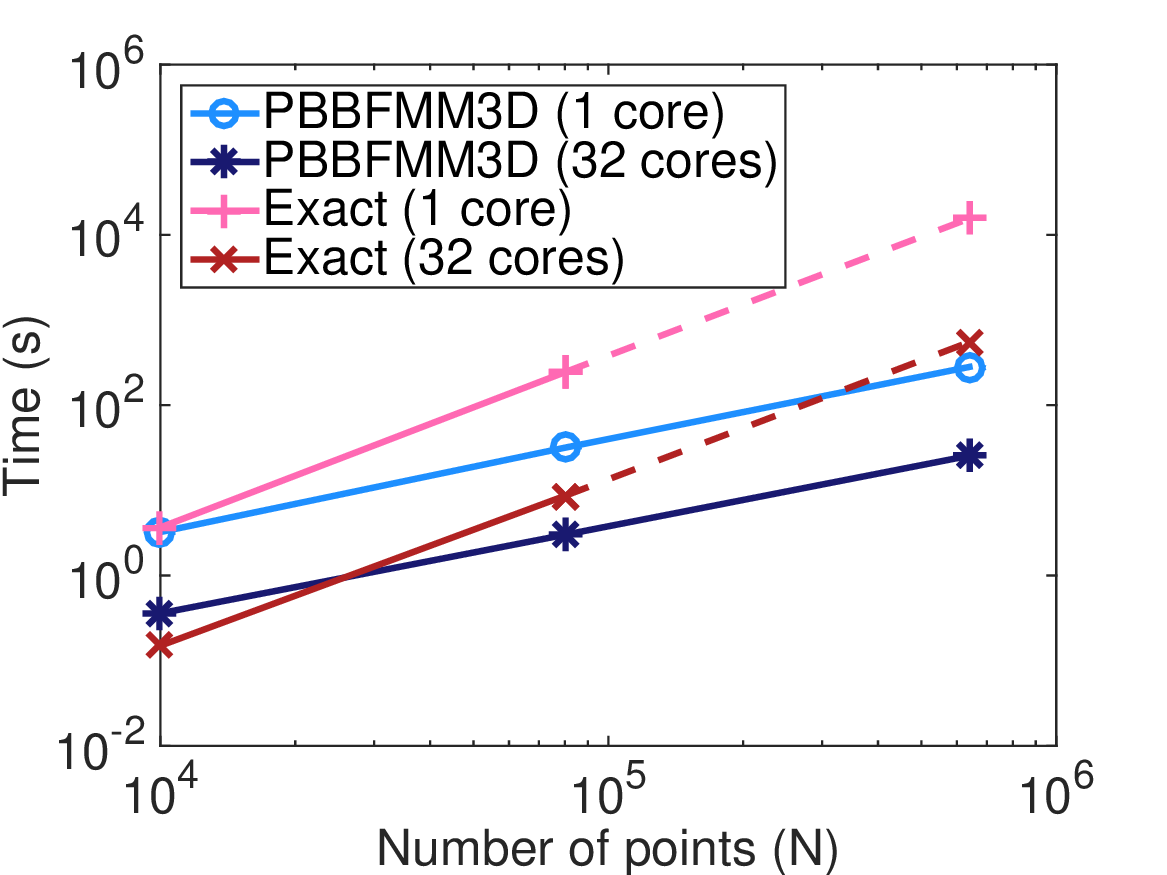}
\caption{Comparison between the original randomized algorithm (Algorithm 5.3 in \cite{halko2011finding}, denoted as `Exact') and the \texttt{PBBFMM3D}-accelerated randomized algorithm. Both methods computed  the top-100 eigen-pairs of the same covariance matrix, where they evaluated 120 matrix-vector products (packed into one matrix-matrix product) twice.
Timing of the original method was not available when $N=64\times 10^4$ because forming the entire covariance matrix requires 3.2 TB of memory.
The error of our accelerated method was at the order of $10^{-4}$.
}
\label{fig:randSVD}
\end{center}
\end{figure}

\cref{fig:randSVD} shows the running time of the original method and our accelerated method. {In Geostatistics applications, the original randomized method with exact matrix-matrix products is the common practice, and therefore we set that as our baseline. This allows us to show the best speedup and scaling one can achieve from leveraging \texttt{PBBFMM3D}. In the original method, we need to evaluate all entries of the covariance matrix and compute matrix-matrix products, both of which scale as $\O(N^2)$. It is worth noting that when $N=64\times 10^4$, forming the entire covariance matrix requires 3.2 TB of memory.}

In the \texttt{PBBFMM3D}-accelerated method, we need to evaluate only $\O(N)$ entries of the covariance matrix. As  \cref{fig:randSVD} shows, our method scales linearly with respect to the problem size $N$, and thus the speedup over the original method  becomes more pronounced when $N$ increases.


\begin{figure}
\begin{center}
\includegraphics[trim=30 60 30 10,clip,width=13.5cm]{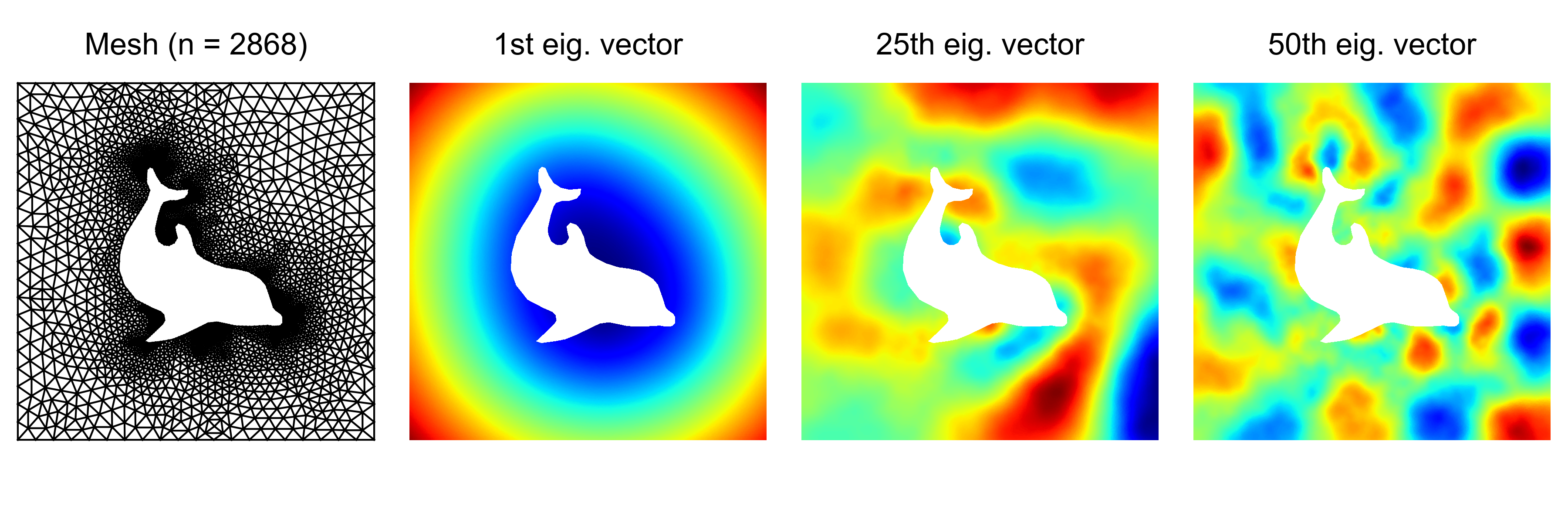}
\caption{An unstructured mesh $N = 693,888$ from \cite{alnaes2015fenics} and three eigenvectors of the covariance matrix with kernel $e^{-r}$. The eigenvectors are  computed via a randomized algorithm~\cite{halko2011finding}, where kernel matrix-vector products are calculated using \texttt{PBBFMM3D}. The left most figure shows the (coarse) mesh with $N=2868$ elements.}
\label{fig:dolfin-eigv}
\end{center}
\end{figure}

{In the last experiment, we  apply \texttt{PBBFMM3D} to  data points lying on an unstructured grid, as shown in \autoref{fig:dolfin-eigv}.} Same as above, we compute the top-50 eigenvectors with the randomized algorithm accelerated by \texttt{PBBFMM3D}. For this experiment, we employ the kernel function $e^{-r}$, and we used $p=5$ Chebyshev nodes in \texttt{PBBFMM3D}.


\section{Conclusions}
\label{sec:conclusion}

We have introduced \texttt{PBBFMM3D}, a black-box algorithm/software for evaluating kernel matrix-vector multiplication on shared-memory machines. The target kernels are non-oscillatory translation invariant functions that are sufficiently smooth away from the origin. The user needs only provide a function routine that returns the kernel value given a source point and a target point, if the kernel function is not already implemented. (The user can also specify the degree of homogeneity and whether the kernel is symmetric or skew-symmetric to active specific optimizations.) Our algorithm requires $\O(N)$ memory and work, where $N$ is the number of data points.

A parallel algorithm is presented in this paper and implemented using \verb|OpenMP| for shared-memory machines. We have presented parallel scalability results on up to 32 cores and achieved at most $19\times$ speedup. We have also presented an application  in geostatistics, where we accelerated the computation of the truncated eigen-decomposition of covariance matrices.

%
%





\bibliographystyle{elsarticle-num} 
\bibliography{pbbfmm}

\begin{thebibliography}{10}
\expandafter\ifx\csname url\endcsname\relax
  \def\url#1{\texttt{#1}}\fi
\expandafter\ifx\csname urlprefix\endcsname\relax\def\urlprefix{URL }\fi
\expandafter\ifx\csname href\endcsname\relax
  \def\href#1#2{#2} \def\path#1{#1}\fi

\bibitem{gray2001n}
A.~G. Gray, A.~W. Moore, N-body'problems in statistical learning, in: Advances
  in neural information processing systems, 2001, pp. 521--527.

\bibitem{hofmann2008kernel}
T.~Hofmann, B.~Sch{\"o}lkopf, A.~J. Smola, Kernel methods in machine learning,
  The annals of statistics (2008) 1171--1220.

\bibitem{ambikasaran2013fast}
S.~Ambikasaran, A.~K. Saibaba, E.~F. Darve, P.~K. Kitanidis, Fast algorithms
  for {B}ayesian inversion, in: Computational Challenges in the Geosciences,
  Springer, 2013, pp. 101--142.

\bibitem{li2014kalman}
J.~Y. Li, S.~Ambikasaran, E.~F. Darve, P.~K. Kitanidis, A {K}alman filter
  powered by {H}2-matrices for quasi-continuous data assimilation problems,
  Water Resources Research 50~(5) (2014) 3734--3749.

\bibitem{greengard1987fast}
L.~Greengard, V.~Rokhlin, A fast algorithm for particle simulations, Journal of
  computational physics 73~(2) (1987) 325--348.

\bibitem{greengard1997new}
L.~Greengard, V.~Rokhlin, A new version of the fast multipole method for the
  {L}aplace equation in three dimensions, Acta numerica 6 (1997) 229--269.

\bibitem{simpson2016acceleration}
R.~Simpson, Z.~Liu, Acceleration of isogeometric boundary element analysis
  through a black-box fast multipole method, Engineering Analysis with Boundary
  Elements 66 (2016) 168--182.

\bibitem{zhao2012fast}
D.~Zhao, J.~Huang, Y.~Xiang, Fast multipole accelerated boundary integral
  equation method for evaluating the stress field associated with dislocations
  in a finite medium, Communications in Computational Physics 12~(1) (2012)
  226--246.

\bibitem{chen2018fast}
C.~Chen, S.~Aubry, T.~Oppelstrup, A.~Arsenlis, E.~Darve, Fast algorithms for
  evaluating the stress field of dislocation lines in anisotropic elastic
  media, Modelling and Simulation in Materials Science and Engineering (2018).

\bibitem{alfke2018nfft}
D.~Alfke, D.~Potts, M.~Stoll, T.~Volkmer, {NFFT} meets {K}rylov methods: Fast
  matrix-vector products for the graph {L}aplacian of fully connected networks,
  Frontiers in Applied Mathematics and Statistics 4 (2018) 61.

\bibitem{ruiz2018nonuniform}
D.~Ruiz-Antolin, A.~Townsend, A nonuniform fast {F}ourier transform based on
  low rank approximation, SIAM Journal on Scientific Computing 40~(1) (2018)
  A529--A547.

\bibitem{fu1998fast}
Y.~Fu, K.~J. Klimkowski, G.~J. Rodin, E.~Berger, J.~C. Browne, J.~K. Singer,
  R.~A. Van De~Geijn, K.~S. Vemaganti, A fast solution method for
  three-dimensional many-particle problems of linear elasticity, International
  Journal for Numerical Methods in Engineering 42~(7) (1998) 1215--1229.

\bibitem{fu2000fast}
Y.~Fu, G.~J. Rodin, Fast solution method for three-dimensional {S}tokesian
  many-particle problems, International Journal for Numerical Methods in
  Biomedical Engineering 16~(2) (2000) 145--149.

\bibitem{greengard2002new}
L.~F. Greengard, J.~Huang, A new version of the fast multipole method for
  screened {C}oulomb interactions in three dimensions, Journal of Computational
  Physics 180~(2) (2002) 642--658.

\bibitem{yoshida2001application}
K.-i. Yoshida, N.~Nishimura, S.~Kobayashi, Application of fast multipole
  {G}alerkin boundary integral equation method to elastostatic crack problems
  in 3d, International Journal for Numerical Methods in Engineering 50~(3)
  (2001) 525--547.

\bibitem{dutt1996fast}
A.~Dutt, M.~Gu, V.~Rokhlin, Fast algorithms for polynomial interpolation,
  integration, and differentiation, SIAM Journal on Numerical Analysis 33~(5)
  (1996) 1689--1711.

\bibitem{gimbutas2003generalized}
Z.~Gimbutas, V.~Rokhlin, A generalized fast multipole method for nonoscillatory
  kernels, SIAM Journal on Scientific Computing 24~(3) (2003) 796--817.

\bibitem{fong2009black}
W.~Fong, E.~Darve, The black-box fast multipole method, Journal of
  Computational Physics 228~(23) (2009) 8712--8725.

\bibitem{ying2004kernel}
L.~Ying, G.~Biros, D.~Zorin, A kernel-independent adaptive fast multipole
  algorithm in two and three dimensions, Journal of Computational Physics
  196~(2) (2004) 591--626.

\bibitem{martinsson2007accelerated}
P.-G. Martinsson, V.~Rokhlin, An accelerated kernel-independent fast multipole
  method in one dimension, SIAM Journal on Scientific Computing 29~(3) (2007)
  1160--1178.

\bibitem{malhotra2015pvfmm}
D.~Malhotra, G.~Biros, {PVFMM}: A parallel kernel independent {FMM} for
  particle and volume potentials, Communications in Computational Physics
  18~(3) (2015) 808--830.

\bibitem{yu2017geometry}
C.~D. Yu, J.~Levitt, S.~Reiz, G.~Biros, Geometry-oblivious {FMM} for
  compressing dense {SPD} matrices, in: Proceedings of the International
  Conference for High Performance Computing, Networking, Storage and Analysis,
  ACM, 2017, p.~53.

\bibitem{lee2018fast}
J.~Lee, A.~Kokkinaki, P.~K. Kitanidis, Fast large-scale joint inversion for
  deep aquifer characterization using pressure and heat tracer measurements,
  Transport in Porous Media 123 (2018) 533--543.

\bibitem{verde2013efficient}
A.~Verde, A.~Ghassemi, et~al., Efficient solution of large-scale displacement
  discontinuity problems using the fast multipole method, in: 47th US Rock
  Mechanics/Geomechanics Symposium, American Rock Mechanics Association, 2013.

\bibitem{verde2015fast}
A.~Verde, A.~Ghassemi, Fast multipole displacement discontinuity method
  ({FM-DDM}) for geomechanics reservoir simulations, International Journal for
  Numerical and Analytical Methods in Geomechanics 39~(18) (2015) 1953--1974.

\bibitem{farmahini2016simulation}
M.~Farmahini-Farahani, A.~Ghassemi, Simulation of micro-seismicity in response
  to injection/production in large-scale fracture networks using the fast
  multipole displacement discontinuity method ({FMDDM}), Engineering Analysis
  with Boundary Elements 71 (2016) 179--189.

\bibitem{coulier2017inverse}
P.~Coulier, H.~Pouransari, E.~Darve, The inverse fast multipole method: using a
  fast approximate direct solver as a preconditioner for dense linear systems,
  SIAM Journal on Scientific Computing 39~(3) (2017) A761--A796.

\bibitem{takahashi2020parallelization}
T.~Takahashi, C.~Chen, E.~Darve, Parallelization of the inverse fast multipole
  method with an application to boundary element method, Computer Physics
  Communications 247 (2020) 106975.

\bibitem{agullo2014task}
E.~Agullo, B.~Bramas, O.~Coulaud, E.~Darve, M.~Messner, T.~Takahashi,
  Task-based {FMM} for multicore architectures, SIAM Journal on Scientific
  Computing 36~(1) (2014) C66--C93.

\bibitem{warren1992astrophysical}
M.~S. Warren, J.~K. Salmon, Astrophysical {N}-body simulations using
  hierarchical tree data structures, SC 92 (1992) 570--576.

\bibitem{warren1993parallel}
M.~S. Warren, J.~K. Salmon, A parallel hashed oct-tree n-body algorithm, in:
  Proceedings of the 1993 ACM/IEEE conference on Supercomputing, 1993, pp.
  12--21.

\bibitem{takahashi2012optimization}
T.~Takahashi, C.~Cecka, E.~Darve, Optimization of the parallel black-box fast
  multipole method on {CUDA}, in: 2012 Innovative Parallel Computing (InPar),
  IEEE, 2012, pp. 1--14.

\bibitem{agullo2016task}
E.~Agullo, B.~Bramas, O.~Coulaud, E.~Darve, M.~Messner, T.~Takahashi,
  Task-based {FMM} for heterogeneous architectures, Concurrency and
  Computation: Practice and Experience 28~(9) (2016) 2608--2629.

\bibitem{trefethen2013approximation}
L.~N. Trefethen, Approximation theory and approximation practice, Siam, 2013.

\bibitem{rasmussen2005gaussian}
C.~E. Rasmussen, C.~K.~I. Williams, Gaussian Processes for Machine Learning
  (Adaptive Computation and Machine Learning), The MIT Press, 2005.

\bibitem{stein1999interpolation}
M.~L. Stein, Interpolation of spatial data : some theory for kriging, Springer
  series in statistics, Springer, New York, 1999.

\bibitem{kitanidis1997introduction}
P.~K. Kitanidis, Introduction to geostatistics: applications in hydrogeology,
  Cambridge University Press, 1997.

\bibitem{oliver2008inverse}
D.~S. Oliver, A.~C. Reynolds, N.~Liu, Inverse theory for petroleum reservoir
  characterization and history matching, Cambridge University Press, 2008.

\bibitem{halko2011finding}
N.~Halko, P.-G. Martinsson, J.~A. Tropp, Finding structure with randomness:
  {P}robabilistic algorithms for constructing approximate matrix
  decompositions, SIAM review 53~(2) (2011) 217--288.

\bibitem{alnaes2015fenics}
M.~S. Aln{\ae}s, J.~Blechta, J.~Hake, A.~Johansson, B.~Kehlet, A.~Logg,
  C.~Richardson, J.~Ring, M.~E. Rognes, G.~N. Wells, The {FEniCS} project
  version 1.5, Archive of Numerical Software 3~(100) (2015).
\newblock \href {https://doi.org/10.11588/ans.2015.100.20553}
  {\path{doi:10.11588/ans.2015.100.20553}}.

\end{thebibliography}

\end{document}